\documentclass[epj]{svjour}

\usepackage{graphics}

\def\gtorder{\mathrel{\raise.3ex\hbox{$>$}\mkern-14mu\lower0.6ex\hbox{$\sim$}}}
\def\ltorder{\mathrel{\raise.3ex\hbox{$<$}\mkern-14mu\lower0.6ex\hbox{$\sim$}}}
\def\etal{{\it et al.}}

\begin{document}

\title{How Well Do We Know the Electromagnetic Form Factors of the Proton?}
\author{J. Arrington\inst{1}}

\institute{Argonne National Lab, Argonne, IL}

\date{Received: date / Revised version: date}

\abstract{Recent measurements of recoil polarization in elastic scattering
have been used to extract the ratio of the electric to the magnetic proton form
factors.  These results disagree with Rosenbluth extractions from cross
section measurements, indicating either an inconsistency between the two techniques,
or a problem with either the polarization transfer or cross section
measurements.  To obtain precise knowledge of the proton form factors, we must
first understand the source of this discrepancy.
} 

\maketitle

\section{Introduction}
\label{intro}

\begin{figure}
\resizebox{0.46\textwidth}{!}{\includegraphics{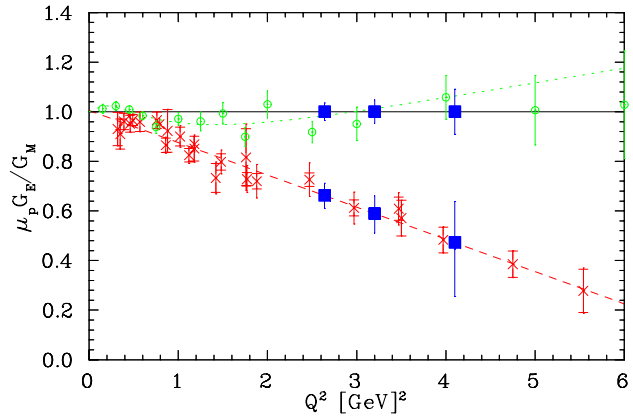} }
\caption{Ratio of $\mu_p G_E/G_M$ from a global
analysis of cross section data~\cite{walker} (circles), and from recent
polarization transfer measurements~\cite{halla} (crosses).  The curves are
the fits~\cite{bosted,halla} to the cross section and polarization data.
The squares are the projected uncertainties for E01-001 (Sec.~\ref{sec:e01001}).}
\label{fig:e01001_proj}
\end{figure}

Elastic form factors measurements probe the charge and magnetization
distributions of the nucleon, and provide strong constraints on models of
nucleon structure.  Prior to the year 2000, all of the high-$Q^2$ proton form
factor data came from cross section measurements, utilizing the
Rosenbluth technique to separate the electric and magnetic
form factors, $G_E$ and $G_M$.  A global analysis of the large body of data on
elastic electron-proton scattering indicated that $G_M$ follows the dipole
form, $G_M = 1/(Q^2+0.71)^2$, with $\sim$5\% deviations (Fig.~\ref{fig:e01001_proj}).
 While the measurements of $G_E$ at high $Q^2$ are significantly less precise,
the extracted ratio of $G_E$ to $G_M$ is roughly constant.

More recent $G_E/G_M$ results, from measurements of the polarization of the
recoil protons, show that $G_E$ falls more rapidly with $Q^2$~\cite{halla} . 
There are significant deviations from the global Rosenbluth analysis above
$Q^2$=1 GeV$^2$, as shown in Fig.~\ref{fig:e01001_proj}.  This discrepancy indicates
that something is wrong with one of these two techniques, or one or more of
the experiments. I will briefly review the two techniques, focusing on
potential sources of systematic uncertainties.  Next, I will give an overview
of a recently completed JLab experiment designed to test the compatibility of
the two techniques.  Finally, I will present an analysis of the previous data
designed to look for possible sources of the discrepancy between the two
techniques. The goal is to determine what kind of errors would have to exist
in the cross section measurements to explain the discrepancies in the form
factors ($\gtorder$100\% error in $G_E$, few \% in $G_M$) and what impact
these errors might have on our knowledge of the proton form factors, as well
as other measurements.

\section{Extractions of the Elastic Form Factors}
\label{sec:comparison}

Rosenbluth extractions (L/T-separations) of the form factors are performed
by measuring elastic electron-proton scattering at fixed virtual photon energy
and momentum ($\nu,\vec q$), while varying the electron energy and scattering angle
to vary the virtual photon polarization, $\epsilon$.  The reduced cross
section, $\sigma_R$, can then be expressed in terms of the form factors, which
depend only on $Q^2$ ($=q^2-\nu^2$):
\begin{equation}
\sigma_R \equiv \frac{d\sigma}{d\Omega} \frac{\epsilon(1+\tau)}{\sigma_{Mott}}
= \tau G_M^2(Q^2) + \epsilon G_E^2(Q^2),
\end{equation}
where $\tau = Q^2/(4 M_p^2)$.  $G_M$ is then extracted from the reduced
cross section at $\epsilon=0$, while $G_E$ is extracted from the
$\epsilon$-dependence.  Due to the $\epsilon / \tau$ weighting of the electric
term relative to the magnetic term, the contribution from $G_E$ decreases as
$1/Q^2$ for a fixed ratio of $G_E/G_M$, and isolating the contribution of
$G_E$ becomes increasingly difficult as $Q^2$ increases.  Because of this, it
is important, and increasingly difficult as $Q^2$ increases, to ensure that the
$\epsilon$-dependent systematics do not overwhelm the uncertainties in the
extraction. Because $\epsilon$ is correlated with beam energy, scattering
angle, and scattered electron energy for a fixed value of $Q^2$, and because
the Mott cross section varies rapidly with angle at fixed $Q^2$, there are
several potential sources of $\epsilon$-dependent errors which might affect
the extracted form factors.

An alternative technique involves using polarized electrons and
measuring the polarization of the recoiling proton.  The ratio of the
transverse to longitudinal components of this polarization is directly
related to $G_E/G_M$.  Measuring a ratio of two polarization components
means that uncertainties in the cross section, beam polarization, and
detector analyzing power all cancel out, significantly reducing the
dominant sources of systematic uncertainty. While this method is clearly
superior at large $Q^2$ values, the discrepancy between the Rosenbluth and
recoil polarization measurements occur at $Q^2$ values as low as $\sim$1
GeV$^2$, where both techniques give precise measurements.

\section{Jefferson Lab Experiment E01-001}
\label{sec:e01001}

Jefferson Lab experiment E01-001 was designed to test the consistency of two
techniques.  In the global analysis of the cross section measurements, a major
concern is the relative normalization of the difference experiments.  While
a normalization factor for each experiment is determined from the best fit to
the entire data sets, there is still room for the normalizations to vary which
could lead to a change in the extracted form factors.  Single experiment
extractions eliminate the effect of normalization uncertainties, which can
dominate the uncertainty in a global analysis. For
the existing single experiment L/T extractions, the dominant sources of
uncertainty come from possible errors that could be correlated with
$\epsilon$.  The largest such uncertainties come from rate dependent
corrections, as the Mott cross section varies rapidly with scattering angle
for fixed $Q^2$, and kinematic-dependent corrections, which may be especially
important for extremely large or small values of the scattered electron momentum,
where effects such as multiple scattering or non-linearities in magnetic
spectrometers may become important.  The goal of E01-001 was to make an
extremely precise Rosenbluth extraction of $G_E/G_M$ in a single measurement,
with careful checks on systematic uncertainties.  Data was taken at three
$Q^2$ values from 2.6 to 4.1 GeV$^2$ to see if the two techniques give
consistent results for the ratio of $G_E/G_M$ in the region where the previous
L/T separations disagree with the new polarization transfer measurements.

E01-001 differs from previous experiments in two main respects: first, we
measured the elastic cross section by detection of the struck proton rather
than the scattered electron, and second, we made simultaneous measurements at
high and low $Q^2$ values for each beam energy.  At fixed $Q^2$, the proton
momentum stays fixed, and so there are no momentum-dependent corrections for
the protons.  In addition, the rate-dependence is dramatically reduced when
detecting protons.  For the kinematics of our experiment, the proton cross
section variation is $<$50\% over the full $\epsilon$-range at each $Q^2$
value.  If the electrons were detected at the corresponding kinematics, the
cross section would vary by 1-2 orders of magnitude between high and low
values of $\epsilon$.  Finally, the cross section is typically a factor of 2-4
less sensitive to uncertainties in beam energy and scattering angles, making
the measurement less sensitive to small uncertainties in the scattering
kinematics.

The main measurement is compared to a normalization point at low $Q^2$
(0.5~GeV$^2$), where the ratio of $G_E/G_M$ is well known and, more
importantly, where the $\epsilon$-range of the measurement is very small
($\Delta\epsilon \approx 0.05$).  Because $\epsilon$ is nearly constant for
the low $Q^2$ measurement, the reduced cross section has a very small, and well
known, $\epsilon$-dependence, and we can use this data as a luminosity monitor
to correct for the beam current, target thickness, and target density
fluctuations.

The experiment was run in May 2002, and data was taken at $Q^2$=2.64, 3.2, and
4.1 GeV$^2$.  The two lower $Q^2$ points will give the most precise results,
and each should provide a better than 7$\sigma$ differentiation
between form factor scaling, $\mu_p G_E/G_M=1$, and the decrease in $G_E/G_M$
measured by the polarization measurements.  The $Q^2=4.1$ point has larger
uncertainties, due to both reduced statistics and increased background, but
should still give a 4$\sigma$ separation between the two results. The existing data
are shown in Fig.~\ref{fig:e01001_proj} along with projected results for
E01-001 under two different assumptions for the extracted ratio. Because the
rate-dependence and kinematic sensitivities are greatly reduced compared to
previous measurements, the errors are dominated by uncertainties which are
uncorrelated at the different $Q^2$ values, making the three measurements
largely independent.  In a typical L/T-separation measurement, any
rate-dependent or momentum-dependent errors would be likely to give similar
effects for the extractions at all $Q^2$ values, and might change the overall
trend of the data for all $Q^2$ points.

\section{Global Reanalysis of Cross Section Data}
\label{sec:reanalysis}

Because of the difficulties in performing L/T-separations at high $Q^2$,
where $G_E$ contributes only a few percent to the cross section, the recoil
polarization technique is more reliable at large $Q^2$.  However, the
disagreement between the two techniques extends to $Q^2 \approx 1$
GeV$^2$.  In this range, the electric form factor contributes 20-30\%
to the cross section at $\epsilon=1$, and Rosenbluth separations give a
precise measurement of $G_E/G_M$.  As mentioned above, the experimental
normalization factors and $\epsilon$-dependent systematics are very important
for these extractions, and it is possible that the disagreement is due to
problems with some subset of the cross section measurements in the global
analysis.  In fact, it has been noted that the results from different
experiments that have extracted $G_E$ are inconsistent at large $Q^2$ values. 
We will first present the individual Rosenbluth separations, and show that the
inconsistencies between different data sets appear to be a combination of
the assumptions in the analyses along with an error in one of the data
sets.  We then will try to determine if the disagreement between the two
techniques can be explained by a simple problem with one or more data sets in
the analysis, or any problems in the analysis itself.

\subsection{Analysis of Individual Rosenbluth Measurements}
\label{sec:single_experiments}

\begin{figure}
\resizebox{0.46\textwidth}{!}{\includegraphics{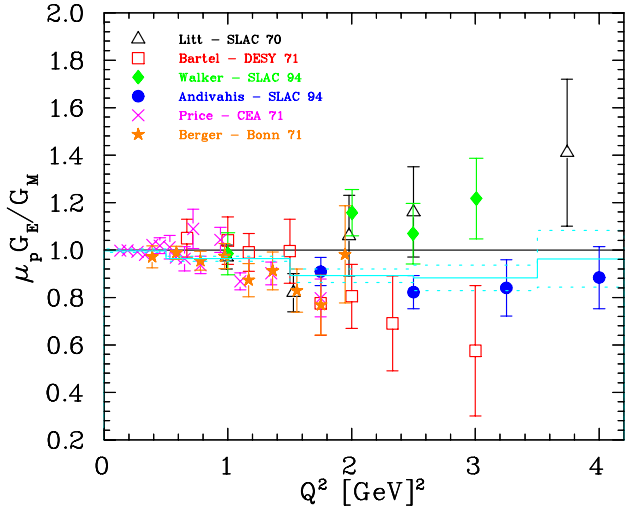} }
\caption{$\mu_p G_E/G_M$ from individual Rosenbluth extractions.}
\label{fig:gegm_lt_raw}
\end{figure}

Figure~\ref{fig:gegm_lt_raw} shows the measurements of $\mu_p G_E/G_M$ for
several different Rosenbluth separation measurements~\cite{walker,single_lt}. 
The data has been binned into five $Q^2$ bins, and the solid and dashed lines
show the weighted average for each bin along with the 1$\sigma$ uncertainties.
 While the combined data set shows approximate form factor scaling, with a
decrease of $\sim$10\% at moderate $Q^2$ values, the individual measurements
have significant scatter about this average.  Comparing each data point to the
average value for its $Q^2$ bin, we get $\chi^2_\nu=1.26$ for 40 degrees of
freedom ({\it i.e.} a confidence level (CL) of 13\%).  The disagreement is
more obvious if we focus on the high $Q^2$ data: $\chi^2_\nu=1.63$ for 17
degrees of freedom for data above $Q^2 = 1.5$ GeV$^2$ (5\% CL).  The scatter
of these results has been used to argue that the experiments are inconsistent,
and that these results should be discarded.


Before concluding that the Rosenbluth extractions are not reliable,
we should examine these data more carefully. There are two problems in this
comparison of `single-experiment' extractions.  First, the Walker data has
a correction at small angles that was discovered by a later SLAC
experiment, but was not taken into account in the original
analysis.  Second, the other extractions shown in Fig.~\ref{fig:gegm_lt_raw}
are not really single-experiment measurements. In three cases (Litt, Price,
and Berger), the values of $G_E$ and $G_M$ come from combining a new set of
cross section measurements with older data at different $\epsilon$ values.
Various procedures have been used to determine a relative normalization
between the two experiments, but the uncertainties from this determination are
either ignored altogether, or applied without the taking into account the fact
that adjusting the normalization of one data set leads to uncertainties that
are highly correlated between the different $Q^2$ values. For the Bartel and
Andivahis data, the form factors are extracted using only the new data, but
that data comes from multiple data sets (taken using different spectrometers,
or detecting protons rather than electrons). For these experiments, a direct
measurement of the relative normalizations at identical kinematics was
possible, and so the normalization factors should be better determined than in
the previous cases, where the normalization factors had to be determined from
data sets that did not have any kinematical overlap. However, while the
normalization factors should be better determined, the correlated nature of the
uncertainties from this determination was not taken into account. Thus, the
extracted form factor ratios shown in Fig.~\ref{fig:gegm_lt_raw} are
not a proper basis for determining the consistency of the different
cross section measurements.

While the ratios shown in Fig.~\ref{fig:gegm_lt_raw} are correct for the
normalizations used in these analyses, the ratio will increase or decrease for
all $Q^2$ values if these normalization factors are varied. Because the ratio
is sensitive to the $\epsilon$-dependence, and because this dependence
decreases with $Q^2$, a shift in the normalization factors would have the
largest effect at higher $Q^2$, and so each of these data sets could be caused
to rise or fall with $Q^2$ by varying the normalization factors. While it may
require a large change in the normalization correction, these analyses can all
be adjusted to reproduce the falloff seen in the polarization transfer
measurement ({\it e.g.} for the most precise measurement, the Andivahis data,
a 4\% change in the measured correction for the low-$\epsilon$ data brings the
ratio into agreement with the Hall A data, but at the cost of introducing a
6$\sigma$ disagreement between the two spectrometers for measurements at
identical kinematics).  In order to study the consistency of the data, we must
avoid the large uncertainties related to the normalization factors.  To do
this, we must either fully take into account the correlated uncertainties
arising from the choice of normalization procedure, or else extract the form
factors from single experiments, where these normalization issues do not arise.
Alternatively, in a global analysis the normalization factors can be better
determined, as the data sets will have significantly more overlap than in the
case where two data sets, one at high $\epsilon$ and the other at low $\epsilon$,
are combined.  This should lead to a more precise determination of the
normalizations.  We can then see if adjustments of the normalization factors
within these uncertainties can significantly change the results.  In the
following sections, we will both analyze the single-experiment Rosenbluth
extractions and perform a global analysis to examine the discrepancy with the
polarization measurements.

\begin{figure}
\resizebox{0.46\textwidth}{!}{\includegraphics{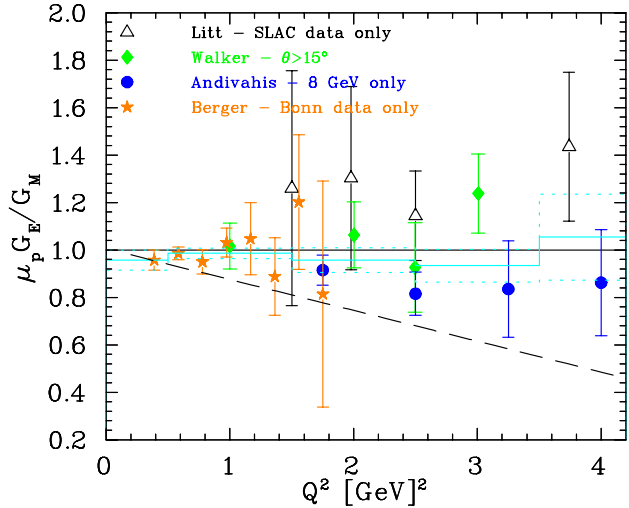} }
\caption{$\mu_p G_E/G_M$ from individual Rosenbluth extractions using only data
from single experiments.  The dashed line is the fit to the Hall A recoil
polarization measurements.}
\label{fig:gegm_lt_safe}
\end{figure}

We begin by repeating the extraction of $G_E/G_M$ for only those experiments
where the $\epsilon$-range was adequate to perform an L/T-separation using the
data from a single detector. For the Walker data, cross sections taken below
15$^\circ$ were excluded to avoid the error from the missing correction.  For
the Andivahis data, we use only the data from the 8 GeV spectrometer, and
exclude the 1.6 GeV spectrometer data (which always provided a single
low-$\epsilon$ point).  Of the other experiments, only the Litt and Berger
data had enough $\epsilon$ range to perform a stand alone Rosenbluth
extraction.  The extractions from this limited data set are shown in
Fig.~\ref{fig:gegm_lt_safe}. The average again yields a ratio that close to
unity, but the data sets are now in better agreement: $\chi^2_\nu=1.18$ for 10
degrees of freedom (30\% CL), for data above $Q^2$=1.5 GeV$^2$.  The average
is clearly well above the polarization data; in fact, all 20 data points lie
above the Hall A fit.

\subsection{Global Fit to Cross Section Data}
\label{sec:globfit}

While these few stand-alone extractions are self-consistent, we would like
to examine the full body of data to determine if the disagreement between the
global analysis and the new polarization results can be at least partially
explained by some problem in the data or analysis.  The fit may be affected by
inclusion of bad data points or data sets.  It may even be that while the {\it
best fit} yields a roughly constant ratio of $G_E$ to $G_M$, adjustment of the
normalization factors for the experiments may bring down this ratio at high
$Q^2$ without significantly decreasing the overall quality of the fit.  To
test such explanations of the discrepancy between the two techniques, a new
global analysis of the cross section measurements is presented which can
be used to test the above possibilities.

\begin{figure}
\resizebox{0.46\textwidth}{!}{\includegraphics{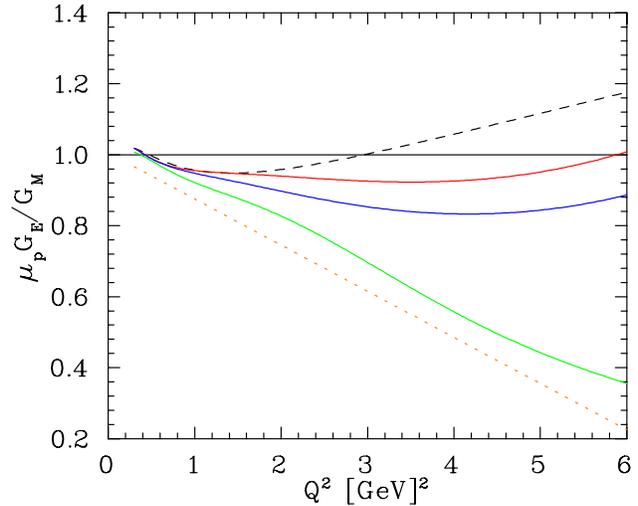} }
\caption{$\mu_p G_E/G_M$ from various global fits: The dotted line is the
fit to the Hall A data, the dashed line is the Bosted fit to the previous
global analysis, the uppermost solid line is the new global fit to the
cross section data, the middle solid line is the fit with three data sets
removed (see text), and the bottom solid line is the combined fit to the
cross section and recoil polarization data.}
\label{fig:fits}
\end{figure}

The global analysis is largely a repeat of the analysis performed in Ref.~\cite{walker}.  We use the same data sets, and perform a combined fit to the
electric and magnetic form factors (with $1/G_E$ and $1/G_M$ parameterized
with 6th order polynomials in $q=\sqrt{Q^2}$) along with a normalization
factor for each of the data sets.  However, there are some differences in the
data sets and fitting procedures.  For the Walker experiment, we exclude the
data below 15 degrees, as discussed in the previous section.  For the Andivahis
measurement, we use the final published cross sections, which were not
available at the time of the previous global analysis.  For the Andivahis and
Berger experiments, we break up the data into subsets, one for each detector
configuration.  Thus, data taken from 13 experiments is broken up into 16
subsets, each with it's own normalization constant.  Finally, we exclude data
below $Q^2=0.3$ GeV$^2$ and above $Q^2 = 10$ GeV$^2$, as we are mainly
interested in the Rosenbluth results in the region where we have polarization
transfer measurements: $0.5 < Q^2 < 6$ GeV$^2$. This initial fit gives similar
results to the Bosted parameterization~\cite{bosted} of the Walker global
analysis.  The top two curves in Fig.~\ref{fig:fits} show the Bosted
fit (dashed line) and the result of the new fit (uppermost solid line).

Removal of the low angle Walker data combined with breaking up the
Andivahis and Bartel data sets slightly reduced the extracted ratio, but it
is still just 5-10\% below scaling, well above the parameterization of the
Hall A data (dotted line). In examining the individual experiments, there are
no data sets that have unacceptably large $\chi^2$ values compared to the
global fit, and no data point that lie beyond 3$\sigma$ from
the fit.  These simple tests do not indicate any clear problems with the data
sets.  However, if a data set has an {\it $\epsilon$-dependent} systematic
uncertainty that is not too large compared to the uncorrelated uncertainties,
then this data set may bias the extracted ratio over a range in $Q^2$ without
having an unusually large $\chi^2$.  In addition, the total $\chi^2$ for the
fit is quite low: $\chi^2$ = 220.4 for 274 degrees of freedom, indicating that
some of the experiments have overestimated the uncertainties. Thus, the simple
statistical test above may not be sufficient, and we need additional checks for
possible bad data sets that could have a large impact on the overall result.

The first test involved repeating the fit 16 times, with a different data set
removed each time.  These fits had only very small changes in the extracted
ratio. The fit was repeated, this time with the removal of the three data
sets whose exclusion caused the largest reduction in the extracted ratio. 
Even this `worst-case' removal of three data sets leads to a small reduction 
($\ltorder$10\%) in the ratio (middle curve in Fig.~\ref{fig:fits}).
Thus, if the discrepancy is caused by errors in the cross section
measurements, it must be a systematic, $\epsilon$-dependent error that impacts
several data sets rather than just a problem with one or two experiments.

Finally, we wish to see if small modifications to the normalization factors 
can remove the inconsistency between the two techniques
without making the overall fit significantly worse.  First, we perform a
constrained fit to the data.  We use the same 16 data sets as in our original
fit, allowing $G_M$ and the normalization factors for each experiment to vary,
but requiring $G_E$ to match the polarization data by constraining the ratio
$\mu_p G_E/G_M=1-0.13(Q^2-0.04)$.  In this way, the normalization factors
will be optimized to reproduce this ratio, as well as maximizing the
consistency between the different data sets. If only small adjustments to the
normalization factors are required, then the $\chi^2$ for this fit should be
only slightly worse than the unconstrained fit. When the ratio
is constrained, the total $\chi^2$ of the fit increases by 60.5 while the
number of degrees of freedom increases by 6 (from 274 to 280). While the total
$\chi^2$ is still close to one, due to the overly conservative error estimates
in some of the data sets, the increase in $\chi^2_\nu$ is 0.20, which is
extremely large for a fit to more than 300 data points. Constraining $G_E/G_M$
to match the Hall A fit clearly gives too much weight to the polarization data,
and so the unconstrained fit was repeated one more time, but this time fitting
to both the cross section data and the recoil polarization data from
Refs.~\cite{halla} (bottom solid line in Fig.~\ref{fig:fits}). Again, the
$\chi^2$ increase is significant: $\chi^2$ increases by 65 while the number of
degrees of freedom increases by 26 (the number of additional data points). In
addition to the fact that the overall fit quality is worse
($\Delta\chi^2_\nu=0.15$ for $\sim$300 degrees of freedom), the recoil
polarization data has larger deviations from the global fit than any of the
other data sets.

From the above tests we conclude that it is not possible to explain the
discrepancy between the two techniques without significant errors in several
data sets, or modifications to the cross section normalization factors that
lead to a significantly worse fit.  This is not too surprising, as the extractions
from single experiments, which do not suffer from uncertainties in the overall
normalization, were in agreement with each other but did not agree with the
recoil polarization measurements.

\section{Conclusions}
\label{sec:conclusions}

The discrepancy between electric form factors extracted from recent
polarization transfer measurements and older cross section data implies either
a fundamental flaw in one of the techniques, or a problem in one or more
experiments.  We have shown that the Rosenbluth extractions of $G_E$ from
previous measurements are self-consistent, providing that one looks only at
analyses that use a single set of data (and after removing the small
angle Walker data). In our new global analysis of the cross section data, we
find no simple explanation for difference between L/T and polarization
measurements.  A problem with the cross section measurement would have to be
an $\epsilon$-dependent error involving several different data sets.

A problem with either the polarization transfer or L/T data might have
significant implications for other experiments using these techniques. If the
problems are shown to be in the cross section data, then the implications are
not limited to other L/T measurements.  To explain the observed
($\gtorder$100\%) inconsistencies in $G_E$, the cross section measurements
must have $\epsilon$-dependent systematic errors on the scale of a few \% or
more (which could be rate-dependent, angle-dependent, or energy-dependent
errors). While recoil polarization measurements can determine the ratio
$G_E/G_M$, the cross sections are still needed to extract the individual form
factors. Errors of a few \% or more in an unknown subset of the cross section
measurements could lead to errors in the form factors at the few \% level, and
could even imply similar errors in the $Q^2$-dependence. In addition, elastic
cross section measurements are often used as a benchmark to determine
normalization for a variety of measurements.

Until we know which result is correct, we cannot be certain of our knowledge of
$G_E$ at $Q^2 > 1$ GeV$^2$.  If E01-001 verifies the polarization transfer
measurements, we must still understand the source of the discrepancy in the
cross section data in order to know the form factors to better than the few \%
level. While it seems likely that such errors would come from cross section
measurements at `extreme' kinematical conditions, and thus would thus have the
largest impact on measurements of the $\epsilon$-dependence, the actual
consequence of the discrepancy between the new and old form factor
measurements will not be clear until we understand why they disagree.

{\it This work is supported in part by the U. S. Department of Energy, Nuclear
Physics Division, under contract W-31-109-ENG-38.}


\begin{thebibliography}{}

\bibitem{walker}
R. C. Walker \etal, Phys. Rev. \textbf{D49}, (1994) 5671.

\bibitem{halla}
M. K. Jones \etal, Phys. Rev. Lett, \textbf{84}, (2000) 1398 ;
O. Gayou \etal, Phys. Rev. \textbf{C64} (2001) 038292 ;
O. Gayou \etal, Phys. Rev. Lett, \textbf{88} (2002) 092301.

\bibitem{bosted} P. E. Bosted, Phys. Rev. {\bf C51}, (1994) 409 .

\bibitem{single_lt}
J. Litt \etal, Phys. Lett. \textbf{31B}, (1970) 40 ;
W. Bartel \etal, Nucl. Phys. \textbf{B58}, (1973) 429 ;
L. Andivahis \etal, Phys. Rev. \textbf{D50}, (1994) 5491 ;
L. E. Price \etal, Phys. Rev. \textbf{D4}, (1971) 45 ;
Ch. Berger \etal, Phys. Lett. \textbf{35B}, (1971) 87.





\end{thebibliography}
\end{document}